\documentclass[amsmath,aps,prd,showpacs,a4paper,10pt]{revtex4}

 \usepackage{epsf}
 \usepackage{graphicx}    

\begin{document}

 \newcommand{\be}[1]{\begin{equation}\label{#1}}
 \newcommand{\ee}{\end{equation}}
 \newcommand{\bea}{\begin{eqnarray}}
 \newcommand{\eea}{\end{eqnarray}}
 \def\disp{\displaystyle}

 \def\gsim{ \lower .75ex \hbox{$\sim$} \llap{\raise .27ex \hbox{$>$}} }
 \def\lsim{ \lower .75ex \hbox{$\sim$} \llap{\raise .27ex \hbox{$<$}} }

 \begin{titlepage}

 \begin{flushright}
 astro-ph/0612746
 \end{flushright}

 \title{\Large \bf Reconstruction of Hessence Dark Energy
 and the Latest Type~Ia Supernovae Gold Dataset}

 \author{Hao~Wei}
 \email[\,email address:\ ]{haowei@mail.tsinghua.edu.cn}
 \affiliation{Department of Physics and Tsinghua Center for
 Astrophysics,\\ Tsinghua University, Beijing 100084, China}

 \author{Ningning~Tang}
 \affiliation{Department of Physics and Tsinghua Center for
 Astrophysics,\\ Tsinghua University, Beijing 100084, China}

 \author{Shuang~Nan~Zhang}
 \affiliation{Department of Physics and Tsinghua Center for
 Astrophysics,\\ Tsinghua University, Beijing 100084, China\\
 Key Laboratory of Particle Astrophysics, Institute of High
 Energy Physics,\\
 Chinese Academy of Sciences, Beijing 100049, China\\
 Physics Department, University of Alabama in Huntsville,
 Huntsville, AL 35899, USA}

 \begin{abstract}\vspace{1cm}
 \centerline{\bf ABSTRACT}\vspace{2mm}
 Recently, many efforts have been made to build dark energy models
 whose equation-of-state parameter can cross the so-called phantom
 divide $w_{de}=-1$. One of them is the so-called hessence dark
 energy model in which the role of dark energy is played by
 a non-canonical complex scalar field. In this work, we develop
 a simple method based on Hubble parameter $H(z)$ to reconstruct
 the hessence dark energy. As examples, we use two familiar
 parameterizations for $H(z)$ and fit them to the latest 182 type
 Ia supernovae Gold dataset. In the reconstruction, measurement
 errors are fully considered.
 \end{abstract}


 \maketitle

 \end{titlepage}

 \renewcommand{\baselinestretch}{1.6}



\section{Introduction}\label{sec1}
Dark energy~\cite{r1} has been one of the most active fields in
 modern cosmology since the discovery of accelerated expansion
 of our universe~\cite{r2,r3,r4,r5,r6,r7,r71}. The simplest candidate
 of dark energy is a tiny positive cosmological constant. However,
 as well-known, it is plagued with the ``cosmological constant
 problem'' and ``coincidence problem''~\cite{r1}. In the
 observational cosmology of dark energy, equation-of-state
 parameter~(EoS) $w_{de}\equiv p_{de}/\rho_{de}$ plays an important
 role, where $p_{de}$ and $\rho_{de}$ are the pressure and energy
 density of dark energy respectively. The most important difference
 between cosmological constant and dynamical scalar fields is that
 the EoS of the former is always a constant, $-1$, while the EoS
 of the latter can be variable during the evolution of the universe.

Recently, evidence for $w_{de}(z)<-1$ at redshift $z<0.2\sim 0.3$
 has been found by fitting observational data
 (see~\cite{r8,r9,r10,r11,r12,r13,r14,r15,r16} for examples).
 In addition, many best-fits of the present value of $w_{de}$ are
 less than $-1$ in various data fittings with different
 parameterizations (see~\cite{r17} for a recent review). The present
 data seem to slightly favor an evolving dark energy with $w_{de}$
 being below $-1$ around present epoch from $w_{de}>-1$ in the near
 past~\cite{r9}. Obviously, the EoS cannot cross the so-called
 phantom divide $w_{de}=-1$ for quintessence or phantom alone.
 Some efforts have been made to build dark energy model whose EoS
 can cross the phantom divide~(see for
 examples~\cite{r9,r18,r19,r20,r21,r22,r23,r24,r25,r26,r27,r28,r65,r66,r67}
 and references therein).

In~\cite{r9}, Feng, Wang and Zhang proposed a so-called quintom
 model which is a hybrid of quintessence and phantom (thus the
 name quintom). Phenomenologically, one may consider a Lagrangian
 density~\cite{r9,r23,r24}
 \be{eq1}
 {\cal L}_{quintom}=\frac{1}{2}\left(\partial_{\mu}\phi_1\right)^2
 -\frac{1}{2}\left(\partial_{\mu}\phi_2\right)^2-V(\phi_1,\phi_2),
 \ee
 where $\phi_1$ and $\phi_2$ are two real scalar fields and play
 the roles of quintessence and phantom respectively. Considering
 a spatially flat Friedmann-Robertson-Walker (FRW) universe and
 assuming the scalar fields $\phi_1$ and $\phi_2$ are homogeneous,
 one obtains the effective pressure and energy density for
 the quintom, i.e.
 \be{eq2}
 p_{quintom}=\frac{1}{2}\dot{\phi}_{1}^2-\frac{1}{2}\dot{\phi}_{2}^2
 -V(\phi_1,\phi_2),~~~~~~~
 \rho_{quintom}=\frac{1}{2}\dot{\phi}_{1}^2
 -\frac{1}{2}\dot{\phi}_{2}^2+V(\phi_1,\phi_2),
 \ee
 respectively. The corresponding effective EoS is given by
 \be{eq3}
 w_{quintom}=\frac{\dot{\phi}_{1}^2-\dot{\phi}_{2}^2
 -2V(\phi_1,\phi_2)}{\dot{\phi}_{1}^2-\dot{\phi}_{2}^2
 +2V(\phi_1,\phi_2)}.
 \ee
 It is easy to see that $w_{quintom}\geq-1$ when
 $\dot{\phi}_{1}^2\geq\dot{\phi}_{2}^2$ while $w_{quintom}<-1$
 when $\dot{\phi}_{1}^2<\dot{\phi}_{2}^2$. The transition occurs
 when $\dot{\phi}_{1}^2=\dot{\phi}_{2}^2$. The cosmological
 evolution of the quintom dark energy was studied in~\cite{r23,r24}.
 Perturbations of the quintom dark energy were investigated
 in~\cite{r29,r30}; and it is found that the quintom model is
 stable when EoS crosses $-1$, in contrast to many dark energy
 models whose EoS can cross the phantom divide~\cite{r28}.

In~\cite{r18}, by a new view of quintom dark energy, one of us
 (H.W.) and his collaborators proposed a novel non-canonical complex
 scalar field, which was named ``hessence'', to play the role of
 quintom. In the hessence model, the phantom-like role is played by
 the so-called internal motion $\dot{\theta}$, where $\theta$ is the
 internal degree of freedom of hessence. The transition from
 $w_{h}>-1$ to $w_{h}<-1$ or vice versa is also possible in the
 hessence model~\cite{r18}. We will briefly review the main points
 of hessence model in Sec.~\ref{sec2}. The cosmological
 evolution of the hessence dark energy was studied in~\cite{r19}
 and then was extended to the more general cases in~\cite{r20}. The
 $w$-$w^\prime$ analysis of hessence dark energy was performed
 in~\cite{r21}.

In this work, we are interested in reconstructing the hessence dark
 energy. In fact, reconstruction of cosmological models is an
 important task of modern cosmology. For instance, the inflaton
 potential reconstruction was extensively studied in~\cite{r31}
 and references therein. The parameterizations and reconstruction
 of quintessence/phantom was considered in~\cite{r32,r33}. The
 other recent reconstructions of quintessence also include
 e.g.~\cite{r34,r35,r36,r37,r38,r39,r40}. The reconstruction of
 k-essence was studied in~\cite{r41,r42,r43}. For the reconstructions
 of other cosmological models, see~\cite{r44,r45,r62,r68,r69} for
 examples. We refer to~\cite{r46} for a recent review on the
 reconstructions of dark energy. In this paper, after a brief review
 of the hessence model, we develop a simple method based on the Hubble
 parameter $H(z)$ to reconstruct the hessence dark energy. As
 examples, we use two familiar parameterizations for $H(z)$
 and fit them to the latest 182 type~Ia supernovae~(SNe~Ia) Gold
 dataset~\cite{r7}. In the reconstruction, measurement errors are
 fully considered.


\section{Hessence dark energy}\label{sec2}
Following~\cite{r18,r19}, we consider a non-canonical complex scalar
 field as the dark energy, namely hessence,
 \be{eq4}
 \Phi=\phi_1+i\phi_2,
 \ee
 with a Lagrangian density
 \be{eq5}
 {\cal L}_h=\frac{1}{4}\left[\,(\partial_\mu \Phi)^2
 +(\partial_\mu\Phi^\ast)^2\,\right]-U(\Phi^2
 +\Phi^{\ast 2})=\frac{1}{2}\left[\,(\partial_\mu \phi)^2
 -\phi^2 (\partial_\mu\theta)^2\,\right]-V(\phi),
 \ee
 where we have introduced two new variables $(\phi,\theta)$ to
 describe the hessence, i.e.
 \be{eq6}
 \phi_1=\phi\cosh\theta,~~~~~~~\phi_2=\phi\sinh\theta,
 \ee
 which are defined by
 \be{eq7}
 \phi^2=\phi_{1}^2-\phi_{2}^2,~~~~~~~\coth\theta=\frac{\phi_1}{\phi_2}.
 \ee
 In fact, it is easy to see that the hessence can be regarded as a
 special case of quintom dark energy in terms of $\phi_1$ and $\phi_2$.
 Considering a spatially flat FRW universe with scale factor $a(t)$
 and assuming $\phi$ and $\theta$ are homogeneous, from Eq.~(\ref{eq5})
 we obtain the equations of motion for $\phi$ and $\theta$,
 \bea
 \ddot{\phi}+3H\dot{\phi}+\phi\dot{\theta}^2+V_{,\phi}=0, \label{eq8}\\
 \phi^2\ddot{\theta}+(2\phi\dot{\phi}+3H\phi^2)\dot{\theta}=0,\label{eq9}
 \eea
 where $H\equiv\dot{a}/a$ is the Hubble parameter, a dot and the subscript
 ``$,\phi$'' denote the derivatives with respect to cosmic time $t$ and
 $\phi$, respectively. The pressure and energy density of the hessence are
 \be{eq10}
 p_h=\frac{1}{2}\left(\dot{\phi}^2-\phi^2\dot{\theta}^2\right)-V(\phi), ~~~~~~~
 \rho_h=\frac{1}{2}\left(\dot{\phi}^2-\phi^2\dot{\theta}^2\right)+V(\phi),
 \ee
 respectively. Eq.~(\ref{eq9}) implies
 \be{eq11}
 Q=a^3 \phi^2\dot{\theta}=const.
 \ee
 which is associated with the total conserved charge within the
 physical volume due to the internal symmetry~\cite{r18,r19}. It turns out
 \be{eq12}
 \dot{\theta}=\frac{Q}{a^3 \phi^2}.
 \ee
 Substituting into Eqs.~(\ref{eq8}) and (\ref{eq10}), they can be
 rewritten as
 \be{eq13}
 \ddot{\phi}+3H\dot{\phi}+\frac{Q^2}{a^6\phi^3}+V_{,\phi}=0,
 \ee
 \be{eq14}
 p_h=\frac{1}{2}\dot{\phi}^2-\frac{Q^2}{2a^6\phi^2}-V(\phi),~~~~~~~
 \rho_h=\frac{1}{2}\dot{\phi}^2-\frac{Q^2}{2a^6 \phi^2}+V(\phi).
 \ee
 It is worth noting that Eq.~(\ref{eq13}) is equivalent to the energy
 conservation equation of hessence, namely,
 $\dot{\rho}_h+3H\left(\rho_h+p_h\right)=0$. The Friedmann equation
 and Raychaudhuri equation are given by
 \be{eq15}
 H^2=\frac{1}{3M_{pl}^2}\left(\rho_h+\rho_m\right),
 \ee
 \be{eq16}
 \dot{H}=-\frac{1}{2M_{pl}^2}\left(\rho_h+\rho_m+p_h\right),
 \ee
 where $\rho_m$ is the energy density of dust matter;
 $M_{pl}\equiv (8\pi G)^{-1/2}$ is the reduced Planck mass. The EoS of
 hessence $w_h\equiv p_h/\rho_h$. It is easy to see that $w_h\geq -1$
 when $\dot{\phi}^2\geq Q^2/(a^6 \phi^2)$, while $w_h<-1$ when
 $\dot{\phi}^2< Q^2/(a^6 \phi^2)$. The transition occurs when
 $\dot{\phi}^2=Q^2/(a^6 \phi^2)$. We refer to the original
 papers~\cite{r18,r19} for more details.


\section{Reconstruction of hessence dark energy}\label{sec3}

Here, we develop a simple reconstruction method based on the Hubble
 parameter $H(z)$ for hessence dark energy. From Eqs.~(\ref{eq15})
 and~(\ref{eq16}), we get
 \be{eq17}
 V(\phi)=3M_{pl}^2H^2+M_{pl}^2\dot{H}-\frac{1}{2}\rho_m,
 \ee
 and
 \be{eq18}
 \dot{\phi}^2-\frac{Q^2}{a^6\phi^2}=-2M_{pl}^2\dot{H}-\rho_m.
 \ee
 Note that
 $$\dot{f}=-(1+z)H\frac{df}{dz}$$
 for any function $f$, where $z=a^{-1}-1$ is the redshift (we set
 $a_0$=1; the subscript ``0'' indicates the present value of the
 corresponding quantity). We can recast Eqs.~(\ref{eq17})
 and~(\ref{eq18}) as
 \be{eq19}
 V(z)=3M_{pl}^2H^2-M_{pl}^2 (1+z)H\frac{dH}{dz}
 -\frac{1}{2}\rho_{m0} (1+z)^3,
 \ee
 \be{eq20}
 \left(\frac{d\phi}{dz}\right)^2-\frac{Q^2}{\phi^2}(1+z)^4 H^{-2}
 =2M_{pl}^2 (1+z)^{-1}H^{-1}\frac{dH}{dz}
 -\rho_{m0}(1+z)H^{-2}.
 \ee
 Introducing the following dimensionless quantities
 \be{eq21}
 \tilde{V}\equiv\frac{V}{M_{pl}^2H_0^2}\,,~~~~~~~
 \tilde{\phi}=\frac{\phi}{M_{pl}}\,,~~~~~~~
 \tilde{H}\equiv\frac{H}{H_0}\,,~~~~~~~
 \tilde{Q}\equiv\frac{Q}{M_{pl}^2H_0}\,,
 \ee
 Eqs.~(\ref{eq19}) and~(\ref{eq20}) can be rewritten as
 \be{eq22}
 \tilde{V}(z)=3\tilde{H}^2-(1+z)\tilde{H}\frac{d\tilde{H}}{dz}
 -\frac{3}{2}\Omega_{m0}(1+z)^3,
 \ee
 \be{eq23}
 \left(\frac{d\tilde{\phi}}{dz}\right)^2
 -\tilde{Q}^2 \tilde{\phi}^{-2}(1+z)^4 \tilde{H}^{-2}=
 2(1+z)^{-1}\tilde{H}\frac{d\tilde{H}}{dz}
 -3\Omega_{m0}(1+z)\tilde{H}^{-2},
 \ee
 where $\Omega_{m0}\equiv\rho_{m0}/(3M_{pl}^2 H_0^2)$ is the present
 fractional energy density of dust matter. Once the $\tilde{H}(z)$,
 or the $H(z)$, is given, we can reconstruct $V(z)$ and $\phi(z)$
 by using Eqs.~(\ref{eq22}) and~(\ref{eq23}) respectively. Then,
 the potential $V(\phi)$ can be reconstructed from $V(z)$ and
 $\phi(z)$ readily. By using Eqs.~(\ref{eq18}) and~(\ref{eq22}),
 we can reconstruct the EoS of hessence
 \be{eq24}
 w_h (z)\equiv\frac{p_h}{\rho_h}
 =\frac{-1+\frac{2}{3}(1+z)\frac{d\ln\tilde{H}}{dz}}
 {1-\Omega_{m0}\tilde{H}^{-2}(1+z)^3}.
 \ee
 The deceleration parameter
 \be{eq25}
 q(z)\equiv-\frac{\ddot{a}}{aH^2}=-1-\frac{\dot{H}}{H^2}
 =-1+(1+z)\tilde{H}^{-1}\frac{d\tilde{H}}{dz}.
 \ee
 After all, it is of interest to reconstruct the kinetic energy
 term of hessence, $K\equiv\dot{\phi}^2/2-Q^2/(2a^6\phi^2)$.
 From Eq.~(\ref{eq18}), we have
 \be{eq26}
 \tilde{K}(z)\equiv\frac{K}{M_{pl}^2H_0^2}
 =(1+z)\tilde{H}\frac{d\tilde{H}}{dz}-\frac{3}{2}\Omega_{m0}(1+z)^3.
 \ee
 It is worth noting that the reconstruction method presented
 here is sufficiently versatile for any $H(z)$.


\section{Examples}\label{sec4}
In this section, as examples, we consider two familiar
 parameterizations for $H(z)$ and fit them to the latest 182 SNe~Ia
 Gold dataset~\cite{r7}. And then, we reconstruct the EoS of hessence
 $w_h(z)$, deceleration parameter $q(z)$, the kinetic energy term of
 hessence $K(z)$, the potential of hessence $V(z)$, and the $\phi(z)$
 as functions of the redshift $z$. Also, we reconstruct the potential
 of hessence as function of $\phi$, namely $V(\phi)$. In our
 reconstruction, measurement errors are fully considered.

\subsection{Parameterizations for $H(z)$ and the latest 182 SNe~Ia Gold dataset}\label{sec4a}

The latest 182 SNe Ia Gold dataset compiled in~\cite{r7} provides
 the apparent magnitude $m(z)$ of the supernovae at peak brightness
 after implementing corrections for galactic extinction,
 K-correction, and light curve width-luminosity correction. The
 resulting apparent magnitude $m(z)$ is related to the luminosity
 distance $d_L(z)$ through (see e.g.~\cite{r58})
 \be{eq27}
 m_{th}(z)=\bar{M}(M,H_0)+5\log_{10} D_L(z),
 \ee
 where
 \be{eq28}
 D_L(z)=(1+z)\int_0^z d\tilde{z}\frac{H_0}{H(\tilde{z};parameters)}
 \ee
 is the Hubble-free luminosity distance $H_0 d_L/c$ in a spatially
 flat FRW universe ($c$ is the speed of light); and
 \be{eq29}
 \bar{M}=M+5\log_{10}\left(\frac{cH_0^{-1}}{\rm Mpc}\right)+25
 =M-5\log_{10}h+42.38
 \ee
 is the magnitude zero offset ($h$ is $H_0$ in units of
 $100~{\rm km/s/Mpc}$); the absolute magnitude $M$ is assumed to be
 constant after the corrections mentioned above. The data points of
 the latest 182 SNe Ia Gold dataset compiled in~\cite{r7} are given
 in terms of the distance modulus
 \be{eq30}
 \mu_{obs}(z_i)\equiv m_{obs}(z_i)-M.
 \ee
 On the other hand, the theoretical distance modulus is defined as
 \be{eq31}
 \mu_{th}(z_i)\equiv m_{th}(z_i)-M=5\log_{10}D_L(z_i)+\mu_0,
 \ee
 where
 \be{eq32}
 \mu_0\equiv 42.38-5\log_{10}h.
 \ee
 The theoretical model parameters are determined by minimizing
 \be{eq33}
 \chi^2(parameters)=\sum\limits_{i=1}^{182}\frac{\left[\mu_{obs}(z_i)
 -\mu_{th}(z_i)\right]^2}{\sigma^2(z_i)},
 \ee
 where $\sigma$ is the corresponding $1\sigma$ error. The parameter
 $\mu_0$ is a nuisance parameter but it is independent of the data
 points. One can perform an uniform marginalization over $\mu_0$.
 However, there is an alternative way. Following~\cite{r58,r59,r64},
 the minimization with respect to $\mu_0$ can be made by expanding
 the $\chi^2$ of Eq.~(\ref{eq33}) with respect to $\mu_0$ as
 \be{eq34}
 \chi^2(parameters)=A-2\mu_0 B+\mu_0^2 C,
 \ee
 where
 $$A(parameters)=\sum\limits_{i=1}^{182}\frac{\left[m_{obs}(z_i)
 -m_{th}(z_i;\mu_0=0,parameters)\right]^2}{\sigma_{m_{obs}}^2(z_i)},$$
 $$B(parameters)=\sum\limits_{i=1}^{182}\frac{m_{obs}(z_i)
 -m_{th}(z_i;\mu_0=0,parameters)}{\sigma_{m_{obs}}^2(z_i)},$$
 $$C=\sum\limits_{i=1}^{182}\frac{1}{\sigma_{m_{obs}}^2(z_i)}.$$
 Eq.~(\ref{eq34}) has a minimum for $\mu_0=B/C$ at
 \be{eq35}
 \tilde{\chi}^2(parameters)=A(parameters)-\frac{B(parameters)^2}{C}.
 \ee
 Therefore, we can instead minimize $\tilde{\chi}^2$ which is
 independent of $\mu_0$, since $\chi^2_{min}=\tilde{\chi}^2_{min}$
 obviously.


 \begin{center}
 \begin{figure}[htbp]
 \centering
 \includegraphics[width=0.6\textwidth]{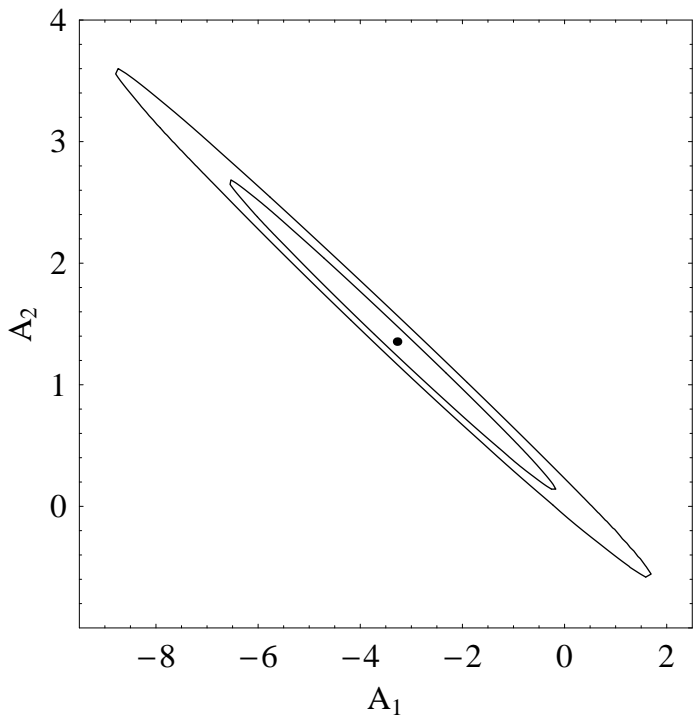}
 \caption{\label{fig1} The $68\%$ and $95\%$ confidence level contours
 in the $A_1$-$A_2$ parameter space for Ansatz~I with the prior
 $\Omega_{m0}=0.30$. The best fit parameters are also indicated
 by a solid point.}
 \end{figure}
 \end{center}


In this work, we consider two familiar parameterizations for
 $H(z)$ and fit them to the latest 182 SNe~Ia Gold data~\cite{r7}.
 At first, we consider the Ansatz~I with
 \be{eq36}
 H(z)=H_0\left[\Omega_{m0}(1+z)^3+A_1 (1+z)+A_2 (1+z)^2
 +\left(1-\Omega_{m0}-A_1-A_2\right)\right]^{1/2},
 \ee
 which has been discussed in~\cite{r41,r47,r12,r15,r62,r68}. Obviously,
 it includes $\Lambda$CDM and XCDM with particular time-independent EoS
 of dark energy as special cases. As shown in~\cite{r47,r41}, even
 for the cases where this ansatz is not exact, one can recover the
 luminosity distance to within $0.5\%$ accuracy using this ansatz in
 the relevant redshift range for the old 157 SNe~Ia Gold
 dataset~\cite{r2}. Therefore, this ansatz is trustworthy to some extent.
 Here, by fitting it to the latest 182 SNe~Ia Gold data~\cite{r7}, for the
 prior $\Omega_{m0}=0.30$~\cite{r72}, we find that the best fit
 parameters~(with $1\sigma$ errors) are $A_1=-3.28\pm 2.11$ and
 $A_2=1.36\pm 0.84$, while $\tilde{\chi}^2_{min}=156.53$ for 180 degrees
 of freedom. The corresponding covariance
 matrix~\cite{r60}~(see also~\cite{r47}) is given by
 \be{eq37}
 Cov(A_1,A_2)=\left(
 \begin{array}{cc} 4.469 & -1.774\\ -1.774 & 0.711 \end{array}
 \right).
 \ee
 In Fig.~\ref{fig1}, we present the corresponding $68\%$ and $95\%$
 confidence level~(c.l.) contours in the $A_1$-$A_2$ parameter space.


 \begin{center}
 \begin{figure}[htbp]
 \centering
 \includegraphics[width=0.6\textwidth]{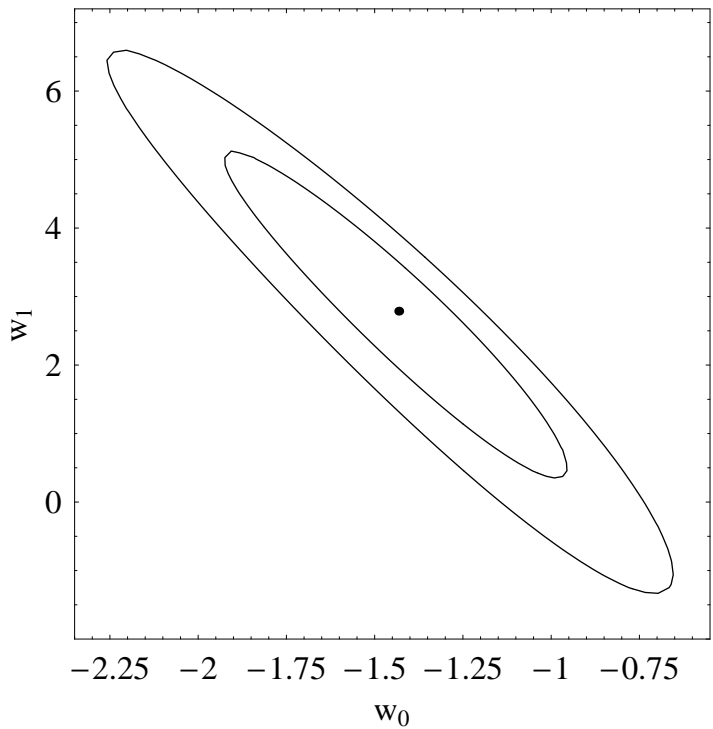}
 \caption{\label{fig2} The $68\%$ and $95\%$ c.l. contours
 in the $w_0$-$w_1$ parameter space for Ansatz~II with the prior
 $\Omega_{m0}=0.30$. The best fit parameters are also indicated
 by a solid point.}
 \end{figure}
 \end{center}


Next, we consider the Ansatz~II with
 \be{eq38}
 H(z)=H_0\left[\Omega_{m0}(1+z)^3
 +\left(1-\Omega_{m0}\right)(1+z)^{3(1+w_0+w_1)}
 \exp\left(-\frac{3w_1 z}{1+z}\right)\right]^{1/2},
 \ee
 which is in fact equivalent to the familiar parameterization
 $w_{de}=w_0+w_1 z/(1+z)$~\cite{r14,r15,r61,r62,r70}. By fitting
 it to the latest 182 SNe~Ia Gold dataset~\cite{r7}, for the prior
 $\Omega_{m0}=0.30$~\cite{r72}, we find that the best fit
 parameters~(with $1\sigma$ errors) are $w_0=-1.43\pm 0.32$ and
 $w_1=2.79\pm 1.55$, while $\tilde{\chi}^2_{min}=156.56$ for 180
 degrees of freedom. The corresponding covariance matrix reads
 \be{eq39}
 Cov(w_0,w_1)=\left(
 \begin{array}{cc} 0.101 & -0.466\\ -0.466 & 2.407 \end{array}
 \right).
 \ee
 In Fig.~\ref{fig2}, we present the corresponding $68\%$ and $95\%$
 c.l. contours in the $w_0$-$w_1$ parameter space.


 \begin{center}
 \begin{figure}[htbp]
 \centering
 \includegraphics[width=0.97\textwidth]{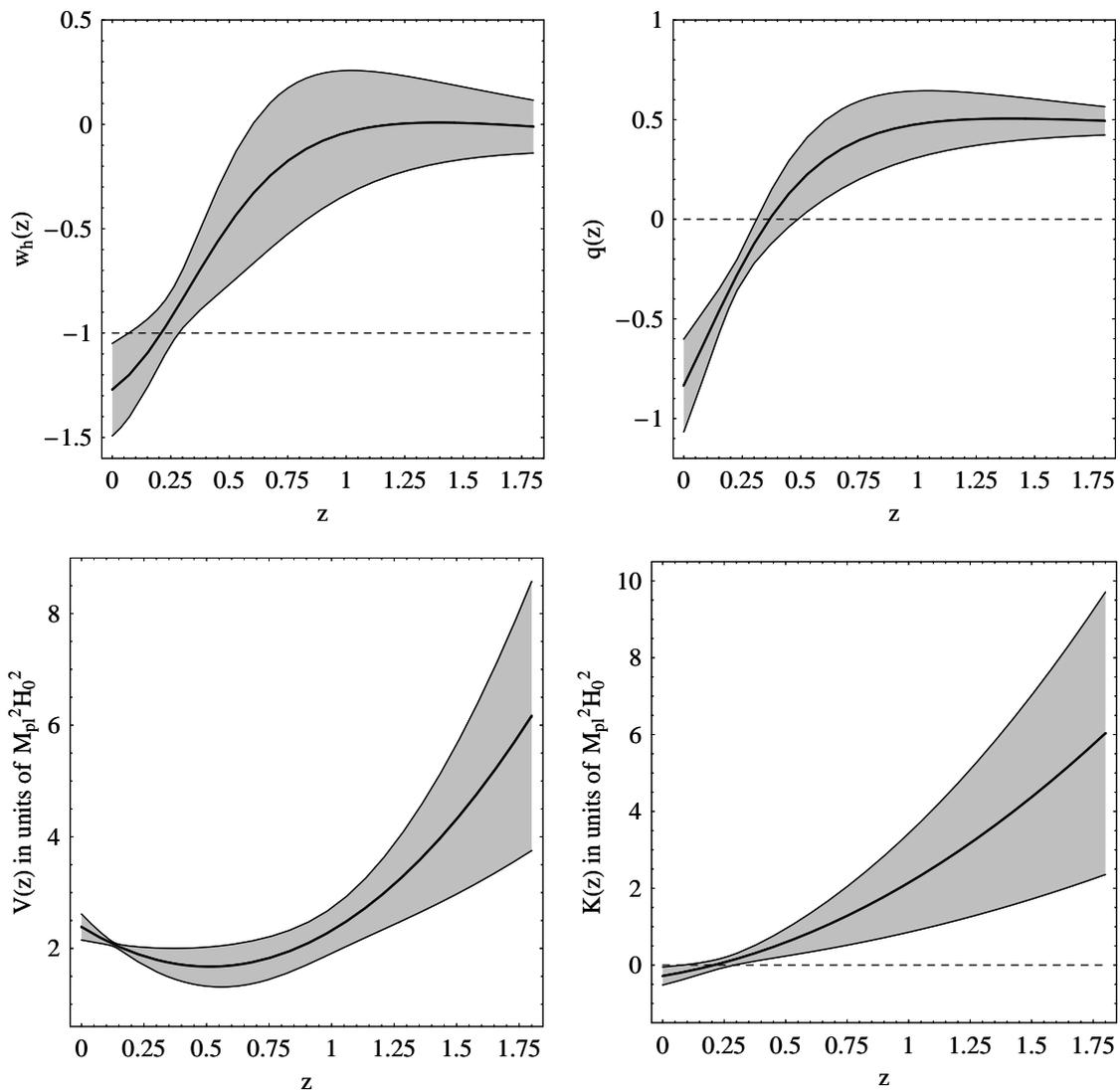}
 \caption{\label{fig3} The reconstructed best fit $w_h(z)$, $q(z)$,
 $V(z)$ and $K(z)$ (thick solid lines) with the corresponding
 $1\sigma$ errors (shaded region) for Ansatz~I.}
 \end{figure}
 \end{center}


\subsection{Reconstruction results}\label{sec4b}

In our reconstruction, measurement errors are fully considered. The
 well-known error propagation equation for any
 $y(x_1,x_2,\ldots,x_n)$,
 \be{eq40}
 \sigma^2(y)=\sum\limits_i^n \left(\frac{\partial y}
 {\partial x_i}\right)^2_{x=\bar{x}}Cov(x_i,x_i)
 +2\sum\limits_{i=1}^n\sum\limits_{j=i+1}^n\left(\frac{\partial y}
 {\partial x_i}\frac{\partial y}{\partial x_j}\right)_{x=\bar{x}}
 Cov(x_i,x_j),
 \ee
 is used extensively (see~\cite{r60} for instance). For Ansatz~I
 with the prior $\Omega_{m0}=0.30$, by using Eqs.~(\ref{eq22}),
 (\ref{eq24})--(\ref{eq26}), (\ref{eq37}), (\ref{eq40}) and the
 corresponding best fit values of $A_1$ and $A_2$, we can
 reconstruct the EoS of hessence $w_h(z)$, deceleration parameter
 $q(z)$, the potential of hessence $V(z)$, and the kinetic energy
 term of hessence $K(z)$ as functions of the redshift $z$, with
 the corresponding $1\sigma$ errors. We show the results in
 Fig.~\ref{fig3}. It is easy to see that $w_h$ crossed $-1$ and
 the universe transited from deceleration ($q>0$) to acceleration
 ($q<0$); the reconstructed $w_h(z)$ is well consistent with the
 three uncorrelated ${\cal W}_{0.25}$, ${\cal W}_{0.70}$ and
 ${\cal W}_{1.35}$ data points with their corresponding $1\sigma$
 error bars for the weak prior~\cite{r7} which are model-independent.

\begin{center}
 \begin{figure}[htbp]
 \centering
 \includegraphics[width=0.97\textwidth]{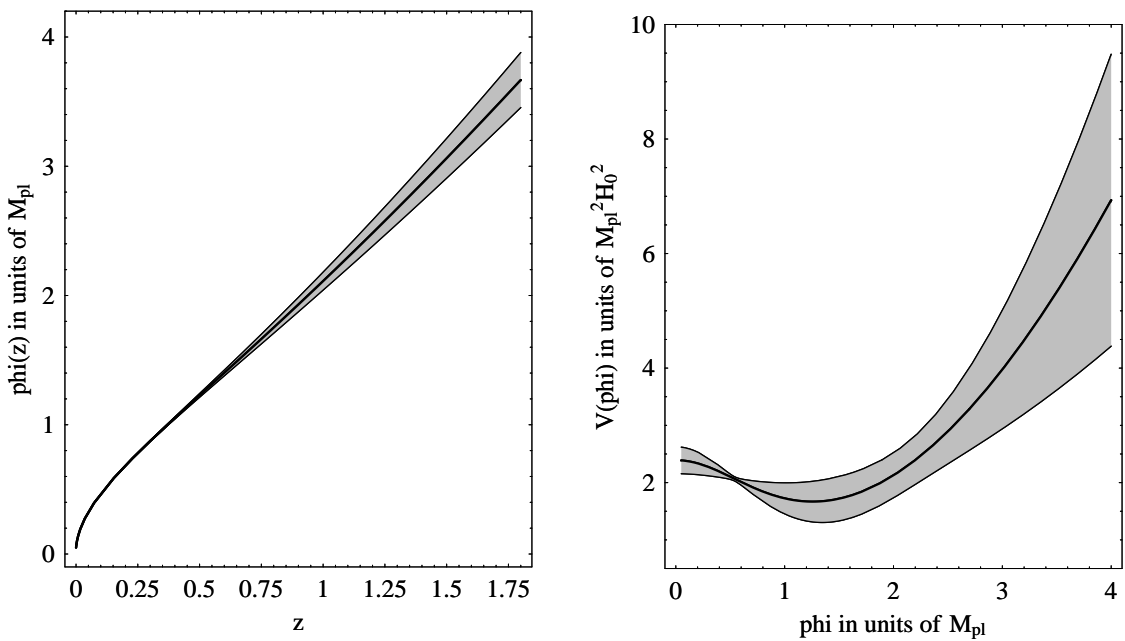}
 \caption{\label{fig4} The reconstructed $\phi(z)$ and $V(\phi)$
 (thick solid lines) with the corresponding $1\sigma$ errors
 (shaded region) for Ansatz~I. See text for details.}
 \end{figure}
 \end{center}

However, the error propagation equation~(\ref{eq40}) is invalid when
 we reconstruct the $\phi(z)$ and hence the $V(\phi)$, since
 $\phi(z)$ is obtained from a differential equation, i.e.
 Eq.~(\ref{eq23}). To evaluate the error propagations, we use
 the Monte Carlo method instead. That is, we generate a multivariate
 Gaussian distribution from the best fit parameters and the
 corresponding covariance matrix. And then, we randomly sample
 $N$ pairs of the parameters $\{A_1,A_2\}$ from this distribution.
 For each pair of $\{A_1,A_2\}$, we can find the corresponding
 $\phi(z)$ and $V(z)$ from Eqs.~(\ref{eq23}) and~(\ref{eq22})
 respectively. Hence, the $V(\phi)$ is in hand. Finally, we can
 determine the mean and the corresponding $1\sigma$ error for the
 $\phi(z)$ and $V(\phi)$ from these $N$ samples, respectively.
 In Fig.~\ref{fig4}, we show the reconstructed $\phi(z)$ and $V(\phi)$
 with the corresponding $1\sigma$ errors for Ansatz~I with the prior
 $\Omega_{m0}=0.30$. In which, we have used the demonstrative initial
 value $\tilde{\phi}_0=0.05$ at $z=0$ and $\tilde{Q}=1$, and have
 chosen the solution with $d\tilde{\phi}/dz>0$ for the reconstructed
 $\phi(z)$; we have done $N=1000$ samplings.


\begin{center}
 \begin{figure}[htbp]
 \centering
 \includegraphics[width=0.97\textwidth]{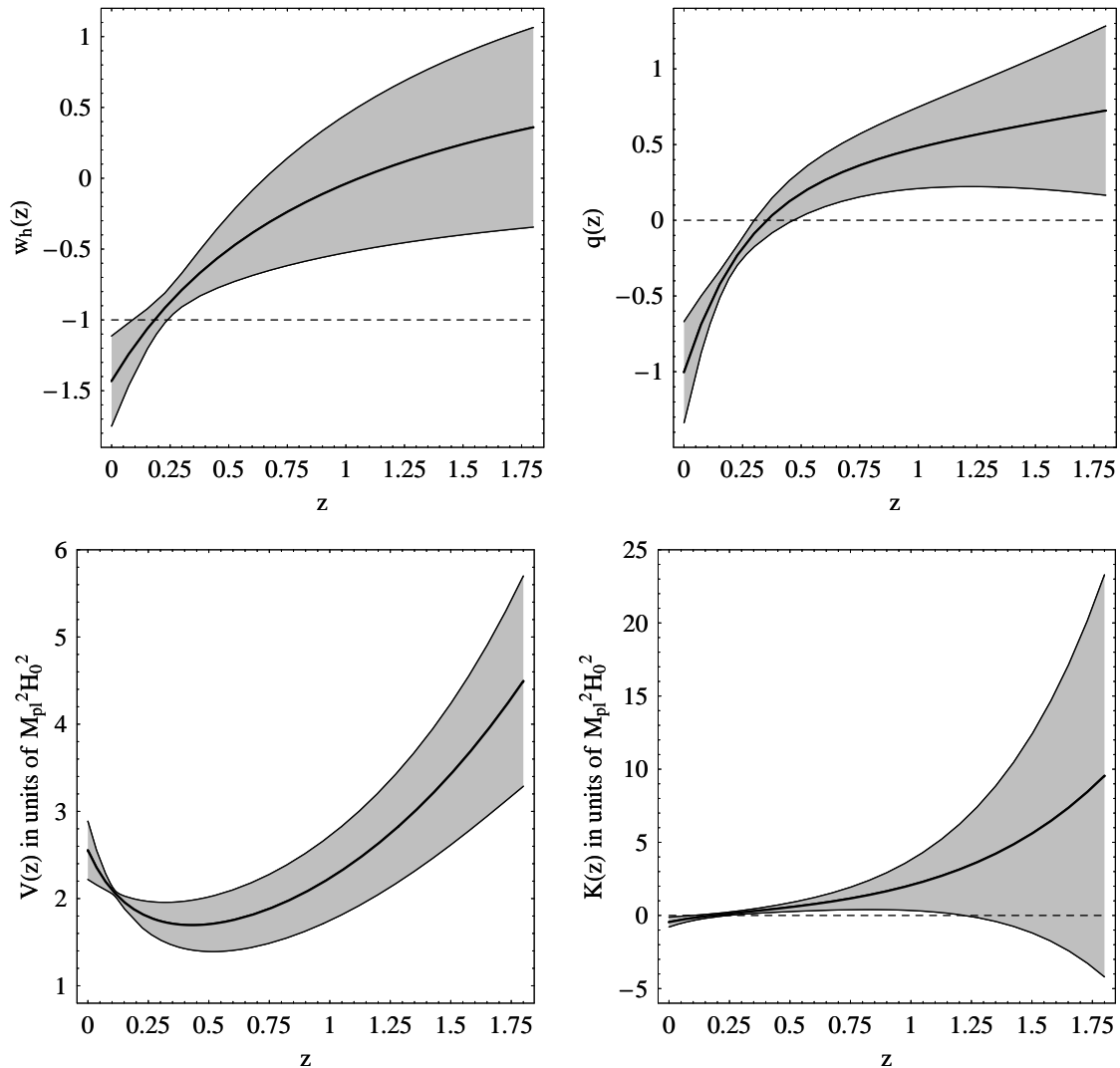}
 \caption{\label{fig5} The reconstructed best fit $w_h(z)$, $q(z)$,
 $V(z)$ and $K(z)$ (thick solid lines) with the corresponding
 $1\sigma$ errors (shaded region) for Ansatz~II.}
 \end{figure}
 \end{center}

\begin{center}
 \begin{figure}[htbp]
 \centering
 \includegraphics[width=0.97\textwidth]{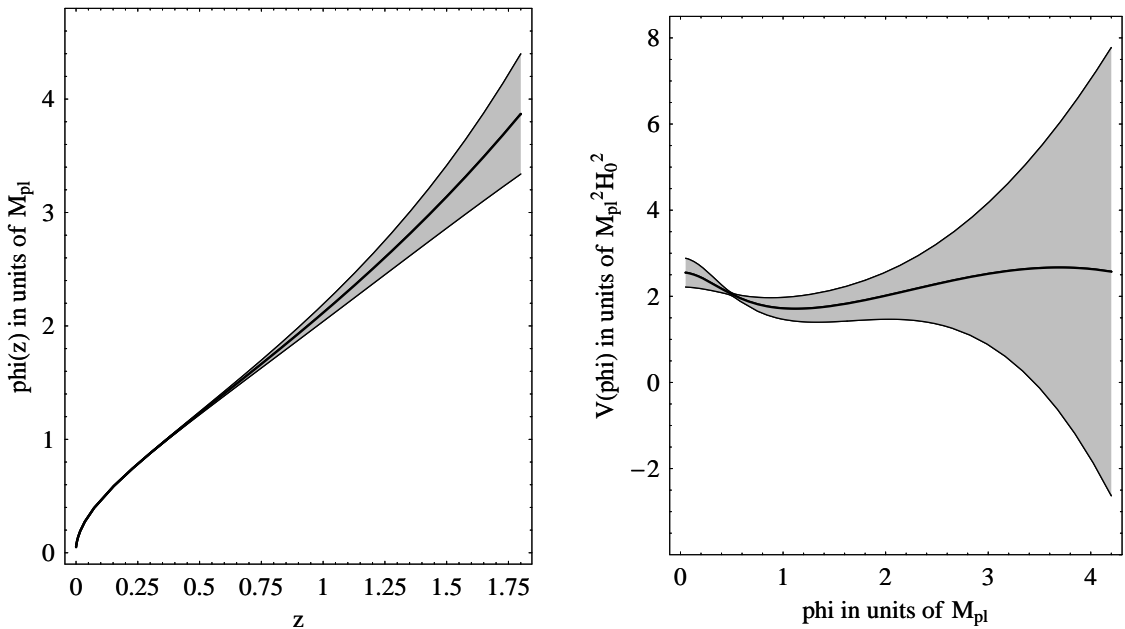}
 \caption{\label{fig6} The reconstructed $\phi(z)$ and $V(\phi)$
 (thick solid lines) with the corresponding $1\sigma$ errors
 (shaded region) for Ansatz~II. See text for details.}
 \end{figure}
 \end{center}

For Ansatz~II, the method to reconstruct the EoS of hessence
 $w_h(z)$, deceleration parameter $q(z)$, the kinetic energy term of
 hessence $K(z)$, the potential of hessence $V(z)$, the $\phi(z)$ as
 functions of the redshift $z$, and the potential of hessence as
 function of $\phi$, namely $V(\phi)$, is the same for Ansatz~I.
 We present the results in Figs.~\ref{fig5} and~\ref{fig6}.
 Once again, it is easy to see that $w_h$ crossed $-1$ and
 the universe transited from deceleration ($q>0$) to acceleration
 ($q<0$); the reconstructed $w_h(z)$ is well consistent with the
 three uncorrelated ${\cal W}_{0.25}$, ${\cal W}_{0.70}$ and
 ${\cal W}_{1.35}$ data points with their corresponding $1\sigma$
 error bars for the weak prior~\cite{r7} which are model-independent.


\section{Discussion}\label{sec5}
In this work, we have developed a simple method based on the Hubble
 parameter $H(z)$ to reconstruct the hessence dark energy. If the
 observational $H(z)$ is obtained, the reconstruction of hessence
 dark energy is ready. It is worth noting that the reconstruction
 method presented here is sufficiently versatile for any $H(z)$.

As examples, we reconstructed the hessence dark energy with two
 familiar parameterizations for $H(z)$. It is easy to see that this
 reconstruction method works well. However, these parameterizations
 for $H(z)$ are not the direct measure of $H(z)$ from observational
 data. We can just say that they are consistent with the recent
 observational data. Can we be in a more comfortable situation?
 In fact, some efforts are aiming to a {\em direct measure} of $H(z)$
 from observational data. Analogous to the estimates of
 $w(z)$~\cite{r8}, a method to obtaining the estimates of $H(z)$ is
 proposed in~\cite{r11} (see also~\cite{r49}). Actually, in the
 latest paper~\cite{r7} by the Supernova Search Team led by Riess,
 a rough estimate of $H(z)$ is obtained by using the new Hubble
 Space Telescope discoveries of SNe Ia at $z\geq 1$. Other new
 method to determine the Hubble parameter as function of redshift,
 $H(z)$, is also proposed in~\cite{r50} recently. In a very different
 way, by using the differential ages of passively evolving galaxies
 determined from the Gemini Deep Deep Survey~(GDDS)~\cite{r51} and
 archival data~\cite{r52}, Simon {\it et al.} determined $H(z)$ in
 the range $0\,\lsim\, z\,\lsim\, 1.8$~\cite{r53,r54,r55,r56,r57,r63}.
 However, up to now, all observational $H(z)$ obtained by various
 methods are too rough to give a reliable reconstruction of
 hessence dark energy. A good news from~\cite{r55} is that a large
 amount of $H(z)$ data is expected to become available in the next
 few years. These include data from the AGN and Galaxy Survey~(AGES)
 and the Atacama Cosmology Telescope~(ACT), and by 2009 an order of
 magnitude increase in $H(z)$ data is anticipated. Therefore, we
 are optimistic to the feasibility of the reconstruction method of
 hessence dark energy proposed in this work.


\section*{ACKNOWLEDGMENTS}
We are grateful to Professor Rong-Gen~Cai for helpful discussions.
 We also thank Zong-Kuan~Guo, Xin~Zhang, Hui~Li, Meng~Su, and Nan~Liang,
 Rong-Jia~Yang, Wei-Ke~Xiao, Jian~Wang, Yuan~Liu, Wei-Ming~Zhang,
 Fu-Yan~Bian for kind help and discussions. We acknowledge partial
 funding support by the Ministry of Education of China, Directional
 Research Project of the Chinese Academy of Sciences and by the
 National Natural Science Foundation of China under project No.~10521001.


\renewcommand{\baselinestretch}{1.1}



\begin{thebibliography}{99}
\bibitem{r72}
The WMAP first year data~(WMAP1) suggest that
 $\Omega_{m0}=0.29\pm 0.07$. Combining the WMAP three year
 data~(WMAP3) with the SNe Gold data, it is found that
 $\Omega_{m0}=0.276^{+0.023}_{-0.031}$, while
 $\Omega_{m0}=0.299^{+0.019}_{-0.025}$ for combining WMAP3 data
 with the CFHTLS lensing data. See Ref.~\cite{r4} for details.
 On the other hand, the SDSS data suggest that
 $\Omega_{m0}=0.299^{+0.037}_{-0.032}$ (see the third one in
 Ref.~\cite{r5}). The SNLS data~\cite{r3} suggest that
 $\Omega_{m0}=0.263\pm 0.042$. The ESSENCE data~\cite{r71}
 give $\Omega_{m0}=0.274^{+0.033}_{-0.020}$ also. So, as in many
 works in the literature, we consider the prior $\Omega_{m0}=0.30$
 throughout this paper.
\end{thebibliography}
\end{document}